\pdfoutput=1

\documentclass[aps,twocolumn,prl,preprintnumbers,superscriptaddress]{revtex4-1}
\usepackage{rotating}
\usepackage[mathscr]{eucal}

\usepackage{amsmath,amssymb,bm,comment}
\usepackage{graphicx}
\usepackage{epsfig}
\usepackage{epstopdf}
\usepackage{hyperref}
\usepackage{longtable}
\usepackage{natbib}
\usepackage{color}

\def\P{{\cal P}}
\def\E{{\cal E}}
\def\N{{\cal N}}
\def\dPdr{{\frac{\partial P}{\partial\rho}}}
\def\dPde{{\frac{\partial P}{\partial\epsilon}}}

\begin{document}

\preprint{CCTP-2012-03}

\title{Towards hydrodynamics without an entropy current}

\author{Kristan Jensen}
\affiliation{Department of Physics, University of Victoria,
Victoria, BC V8W 3P6, Canada}
\author{Matthias Kaminski}
\affiliation{Department of Physics, University of Washington, Seattle, WA 98195, USA}
\author{Pavel Kovtun}
\affiliation{Department of Physics, University of Victoria,
Victoria, BC V8W 3P6, Canada}
\author{Ren\'e Meyer}
\affiliation{Crete Center for Theoretical Physics, Department of Physics, University of Crete, 71003 Heraklion, Greece}
\author{Adam Ritz}
\affiliation{Department of Physics, University of Victoria,
Victoria, BC V8W 3P6, Canada}
\author{Amos Yarom}
\affiliation{Department of Physics, Technion, Haifa 32000, Israel}
\date{March 15, 2012}
\begin{abstract}
\noindent
We present a generating functional which describes the equilibrium thermodynamic response of a relativistic system to external sources. A variational principle gives rise to constraints on the response parameters of relativistic hydrodynamics without making use of an entropy current.  Our method reproduces and extends results available in the literature. It also provides a technique for efficiently computing $n$-point zero-frequency hydrodynamic correlation functions without the need to solve the equations of hydrodynamics.
\end{abstract}
\maketitle

\emph{Introduction.}---Hydrodynamics is a generic effective theory, valid on distance scales much longer than the typical mean free path, and applicable to many diverse physical theories at finite temperature \cite{LL6}. The equations of hydrodynamics are characterized by several response parameters which need to be specified for each particular system. 
These response parameters are usually constrained by a set of equalities and inequalities which are conventionally determined by requiring the existence of a local entropy current with positive semi-definite divergence.
In this Letter, we will systematically derive the equality-type constraints on the response parameters using a variational principle.

In the hydrodynamic regime, a relativistic system can be described in terms of a velocity field $u^{\mu}$, normalized such that $u_{\mu}u^{\mu}=-1$, and a temperature $T$. When there is a conserved $U(1)$  charge, the corresponding chemical potential $\mu$ provides an additional hydrodynamic degree of freedom. If the $U(1)$ symmetry is spontaneously broken, the emerging Goldstone boson $\phi$ also turns into a hydrodynamic degree of freedom. 

The energy-momentum tensor $T^{\mu\nu}$ and the (non-anomalous) charge current $J^{\mu}$ may be expressed through constitutive relations in terms of the hydrodynamic variables and their gradients. 
The kinematic equations for hydrodynamics then amount to energy-momentum and charge conservation,
\begin{align}
\label{E:conservation}
	D_{\mu}T^{\mu\nu} &= F^{\nu\rho}J_{\rho}, \qquad D_{\mu}J^{\mu}  = 0.
\end{align}
In \eqref{E:conservation} $F^{\mu\nu}$  is the field strength of a background gauge field $A_{\mu}$ conjugate to $J^{\mu}$. 
The covariant derivative $D_{\mu}$ depends on a background metric $g_{\mu\nu}$.

In $d$ spacetime dimensions, we decompose the energy-momentum tensor and $U(1)$ current into scalars, vectors and tensors of the $SO(d{-}1) \subset SO(d{-}1,1)$ symmetry preserved by $u^{\mu}$,
\begin{align}
\begin{split}
\label{E:TJdecomp}
	T^{\mu\nu} &= \E u^{\mu}u^{\nu} + \P \Delta^{\mu\nu} + q^{\mu}u^{\nu}+q^{\nu}u^{\mu} + \tau^{\mu\nu}, \\
J^{\mu} &= \N u^{\mu}+\nu^{\mu},
\end{split}
\end{align}
where $\Delta^{\mu\nu} = g^{\mu\nu} + u^{\mu}u^{\nu}$
is a projection matrix, $q^{\mu}u_{\mu}=\nu^{\mu}u_{\mu} = 0$ and $u_{\mu}\tau^{\mu\nu}=g_{\mu\nu}\tau^{\mu\nu}=0$. The scalars $\mathcal{E}$, $\mathcal{P}$, and $\N$ along with the vectors $q^{\mu}$, $\nu^{\mu}$ and the tensor $\tau^{\mu\nu}$ may be written as local functions of the hydrodynamic variables and their derivatives. 
In the hydrodynamic approximation, the constitutive relations ~(\ref{E:TJdecomp})
can be written in a derivative expansion \cite{Bhattacharyya:2008jc,*Baier:2007ix}.

Several considerations come into play in determining which tensor structures can contribute to the quantities in~\eqref{E:TJdecomp}. 
First, we note that there is an inherent ambiguity in defining the velocity field, temperature, and chemical potential out of equilibrium. We may always redefine $T \to T+\delta T$, $\mu\to\mu+\delta\mu$, and $u^{\mu} \to u^{\mu} + \delta u^{\mu}$ such that $\delta T$, $\delta \mu$, and $\delta u^{\mu}$ vanish in the absence of gradients. Such field redefinitions are called changes of frame. A canonical choice of frame is the Landau frame in which $q^{\mu} = 0$, $\mathcal{E}= \epsilon$, and $\N = \rho$ with $\epsilon$ and $\rho$ the energy and charge densities in the absence of gradients. Even after choosing a frame not all tensor structures are allowed; as it turns out, the existence of an entropy current together with the Onsager relations leads to restrictions on the allowed tensor structures \cite{LL6}. 

The restrictions imposed by the existence of an entropy current are either inequalities or equalities. For example, in the presence of an electric field, the tensor decomposition of the current, $J^i = \kappa \Delta^{i\mu}\partial_{\mu} \frac{\mu}{T} +\sigma E^i$, is subject to an equality-type relation, $\kappa = -\sigma T$. 
In this Letter, we systematically show how relations of this type are enforced by the equilibrium properties of the theory in the presence of external sources $A_\mu$ and $g_{\mu\nu}$, and follow from a variational principle. As a result, we learn that the hydrodynamic constitutive relations are constrained both by symmetry and by the
need to consistently describe static equilibria with external sources.

\emph{Equilibria.}---A time-independent equilibrium configuration can be characterized by a constant timelike vector $V^{\mu}$, where $V^\mu = (1, {\bf 0})$ in suitable coordinates. Starting from a source-free equilibrium configuration we assume that finite sources can be turned on adiabatically while maintaining equilibrium. In other words, we will only be considering configurations in which the Lie derivative with respect to $V^{\mu}$, $\mathcal{L}_V$, vanishes when acting on thermodynamic quantities or sources. Furthermore, we will be studying configurations with a finite static correlation length. Thus, Euclidean correlation functions fall off exponentially at large distances, implying that zero-frequency Fourier-space correlators are analytic at low momentum. 

Correlators in the equilibrium configuration can be obtained by differentiating a generating functional with respect to the sources. Indeed, consider the set of all zero-frequency correlation functions, expanded to $m^{\rm th}$ order in momenta about zero. We call these $n$-point functions truncated correlators. After a Fourier transform, we obtain approximate position-space correlation functions valid on length scales much larger than the correlation length of the system, much like a multipole approximation characterizes a localized distribution on large scales. 
Integrating the truncated functions over sources leads to one-point functions, which will be local functions of the sources. These may be further integrated to obtain the equilibrium generating functional for truncated correlation functions
\begin{equation}
\label{E:Wgeneral}
W_m = \int \!\!d^{d}x\,\mathcal{L}[\text{sources}(x)],
\end{equation}
where $\mathcal{L}$ includes terms with up to $m$ derivatives \footnote{There exist several distinct proposals in the literature which define a
dynamical effective action for non-dissipative 
hydrodynamics \cite{Schutz:1970my,*Dixon:1979,*Brown:1992kc,*Son:2002zn,*Jackiw:2004nm,*Dubovsky:2005xd,*Dubovsky:2011sj}.}.

In order for $W_m$ to be diffeomorphism and gauge invariant, $\mathcal{L}$ must be constructed from local diffeomorphism and gauge invariant scalars, possibly in combination with $V^{\mu}$. In addition, $\mathcal{L}$ can depend on observables that are local in space but non-local in Euclidean time such as the invariant length of the time circle in the Euclideanized theory  $L$, and the Polyakov loops $P_A$ of any $U(1)$ gauge fields. 
Since $\mathcal{L}_V=0$, we find $L = \beta \sqrt{-V^2}$ 
and $\ln P_A=\beta V^{\mu}A_{\mu}$, where $\beta$ is the coordinate periodicity of the time circle \footnote{Some readers may be familiar with the Euclidean expression $L= \int_0^\beta d\tau \sqrt{g_{\tau\tau}}$ which can be obtained from $L=\beta \sqrt{-V^2}$ by setting $V^\mu = (1,{\bf 0})$ and rotating to Euclidean time.
}.
We identify the temperature $T$, the chemical potential $\mu $, and the velocity field $u^{\mu}$ as
\begin{equation}
\label{E:EquilibriumDef}
	T=1/L,
	\qquad
	\mu = \ln P_A/L,
	\qquad
	u^{\mu}=\frac{V^{\mu}}{\sqrt{-V^2}}.
\end{equation}
The parameters $T$, $\mu$, and $u^{\mu}$ depend on position through
 $A_{\mu}$ and $V_{\mu}$.

Suppose that there are $N_n$ scalar quantities at $n^{th}$ order in a derivative expansion. We will denote them by $s_{n,1},\,,s_{n,2},\ldots,\,s_{n,N_n}$. For instance, in a theory containing a single conserved current (corresponding to an unbroken symmetry) we have $s_{0,1} = T$ and $s_{0,2}=\mu$. The most general generating functional for truncated zero-frequency correlators is of the form
\begin{equation}
\label{E:Wderivative}
W_m = \int d^{d}x\,\sqrt{-g}\left[ P(s_{0})+\sum_{n=1}^m \sum_{i=1}^{N_n}\alpha_{n,i}(s_{0})s_{n,i}\right],
\end{equation}
where the $\alpha_{n,i}$ and $P$ are functions of the zeroth order scalars which we denoted collectively by $s_0$. In the source-free equilibrium state, all of the derivative contributions to~\eqref{E:Wderivative} vanish, so that $W_m$ is the logarithm of the exact equilibrium partition function. Thus, we identify $P$ with the pressure of the source-free equilibrium state. 

We obtain one-point functions of the energy-momentum tensor and conserved current by varying $W_m$ with respect to the metric and gauge field,
\begin{equation}
\label{E:TJ}
\langle T^{\mu\nu} \rangle = \frac{2}{\sqrt{-g}}\frac{\delta W_m}{\delta g_{\mu\nu}}, \qquad \langle J^{\mu}\rangle = \frac{1}{\sqrt{-g}}\frac{\delta W_m}{\delta A_{\mu}}.
\end{equation}
If we denote the set of $n^{\rm th}$ order transverse vectors and transverse traceless tensors by $v_{n,i}$ and $t_{n,i}$ then, on comparing \eqref{E:TJ} to \eqref{E:TJdecomp}, we find 
\begin{align}
\notag
	\E &=\sum_{n=0}^m \sum_{i=1}^{N_n} \epsilon_{n,i}s_{n,i}, & \P &=\sum_{n=0}^m \sum_{i=1}^{N_n} \pi_{n,i} s_{n,i}, \\
\label{E:expansion}
	\N & =\sum_{n=0}^m \sum_{i=1}^{N_n} \phi_{n,i}s_{n,i}, & q^{\mu}&=\sum_{n=0}^m \sum_{i=1}^{N_n} \gamma_{n,i}v^{\mu}_{n,i}, \\
\notag
 \nu^{\mu}&= \sum_{n=0}^m \sum_{i=1}^{N_n} \delta_{n,i} v^{\mu}_{n,i}, & \tau^{\mu\nu} &= \sum_{n=0}^m \sum_{i=1}^{N_n} \theta_{n,i} t_{n,i}^{\mu\nu},
\end{align}
where the $\epsilon$'s, $\pi$'s, $\phi$'s, $\gamma$'s, $\delta$'s and $\theta$'s are determined in terms of the $\alpha$'s and their derivatives. While the most general expression for the energy-momentum tensor and current takes the form \eqref{E:TJdecomp}, the fact that $T^{\mu\nu}$ and $J^{\mu}$ have been obtained from a local generating functional implies that not all tensor, vector and scalar structures are allowed, and that relations of the form \eqref{E:expansion} hold. 

The  gauge and diffeomorphism invariance of $W_m$ ensures that the solution  \eqref{E:expansion} provided by \eqref{E:TJ} will satisfy the hydrodynamic equations \eqref{E:conservation}. Matching the thermodynamic theory to the effective hydrodynamic description~\eqref{E:expansion} gives the constitutive relations in a particular frame, which we call the thermodynamic frame. In this frame, the values for the temperature, chemical potential and velocity field remain unchanged from their equilibrium definitions \eqref{E:EquilibriumDef} after the hydrodynamic equations have been solved.

In what follows we will give several explicit examples of systems where the relations \eqref{E:expansion} are obtained from the generating functional \eqref{E:Wderivative}. Some of these systems have been analyzed in the literature by requiring the existence of an entropy current.  In all our examples, the non-dissipative constraints obtained using the entropy current method match those obtained here. 

\emph{Example 1: Ideal superfluids.}---We begin by constructing the generating functional and computing the one-point functions for a superfluid to zeroth order in derivatives (i.e. an ideal superfluid, see \cite{Son:2000ht,Herzog:2008he} for a brief review). In addition to the zeroth order scalars $s_{0,1}=T$ and $s_{0,2}=\mu$ we can, a priori, construct two scalars from the extra hydrodynamic degree of freedom associated with the Goldstone boson, $\xi^{\mu} \xi_{\mu} = -\xi^2$ and $u^{\mu}\xi_{\mu}$, where $\xi^{\mu}$ is the gauge invariant combination $\xi_{\mu} = -\partial_{\mu}\phi + A_{\mu}$. Since $\mathcal{L}_V\phi=0$ implies that $u^{\mu}\xi_{\mu}=\mu$, only $s_{0,3} = \xi^2$ is an independent scalar. According to \eqref{E:Wderivative}, the generating functional takes the form
\begin{equation}
W_0 = \int d^{d}x\sqrt{-g} P(T,\mu,\xi^2).
\end{equation}
Using~\eqref{E:TJ} we find
\begin{align}
\begin{split}
\label{E:TJsuperfluid}
\langle T^{\mu\nu}\rangle &= \epsilon u^{\mu}u^{\nu}+P\Delta^{\mu\nu} + f \xi^{\mu}\xi^{\nu}, \\
\langle J^{\mu}\rangle & = \rho u^{\mu}  - f \xi^{\mu}, \qquad u^{\mu}\xi_{\mu}=\mu,
\end{split}
\end{align}
with $dP = s dT + \rho d\mu + \frac{1}{2}fd\xi^2$ and $ \epsilon = T s + \mu \rho - P$ \footnote{For an ideal superfluid, a sufficient condition for the conservation laws (\ref{E:conservation}) to hold in equilibrium is that $D_\mu (f\xi^\mu)=0$, the equation of motion for $\phi$. It would be interesting to identify the geometric criteria which will determine equilibrium configurations of superfluids in the presence of an external
metric and gauge field. We thank S. Bhattacharyya for emphasizing this point.}. 
The expressions in~\eqref{E:TJsuperfluid} 
precisely match those of an ideal superfluid in the notation of~\cite{Bhattacharya:2011tra}. For $\xi=0$ we recover the standard expression for an ideal normal fluid. 

\emph{Example 2: Parity-violating fluids.}---For parity-violating theories in $2{+}1$ dimensions with a conserved $U(1)$ charge, the zeroth order scalars are $s_{0,1} = T$ and $s_{0,2} = \mu$. At first order in the derivative expansion there are a priori three scalars, $D_{\mu}u^{\mu}$, $u^{\mu}\partial_{\mu} T$, and $u^{\mu}\partial_{\mu}\mu$. However,  all three scalars vanish identically since $\mathcal{L}_V=0$. There are two non-vanishing pseudo-scalars, $\tilde{s}_{1}$ and $\tilde{s}_{2}$ defined in Table \ref{T:PVbasis} which specify the magnetic field and vorticity.
Thus, we have the generating functional 
\begin{equation}
\label{E:WPV1}
W_1 = \int d^{3}x\sqrt{-g} \left[  P(T,\mu) + \tilde{\alpha}_{1}\tilde{s}_{1}+\tilde{\alpha}_2\tilde{s}_2  \right]\,.
\end{equation}
The notation in~\eqref{E:WPV1} deviates slightly from~\eqref{E:Wderivative} in that the coefficients of parity odd tensors are adorned with a tilde. Furthermore, since all the results in this example involve tensors with one derivative, we have also dropped the derivative index $n$. We use the same simplified notation for the quantities in~\eqref{E:expansion}. 

An exhaustive list of all possible first derivative tensors can be found in~\cite{Jensen:2011xb}. The constraint $\mathcal{L}_V=0$ leads to the restricted list given in Table~\ref{T:PVbasis}. Indeed, in equilibrium we find that
\begin{align}
\begin{split}
\partial_{\mu}T &= - T a_{\mu}, \quad \partial_{\mu} \mu = - \mu a_{\mu} + E_{\mu}, \\
\label{E:Du}
D_{\mu}u_{\nu} &= -u_{\mu}a_{\nu}+\omega_{\mu\nu},
\end{split}
\end{align}
are satisfied identically, where we have defined
$a^{\mu}=u^{\nu}D_{\nu}u^{\mu}$, and $\omega^{\mu\nu}=\frac{\Delta^{\mu\rho}\Delta^{\nu\sigma}}{2}(D_{\rho}u_{\sigma}-D_{\sigma}u_{\rho})$. Thus, the shear tensor $\sigma^{\mu\nu}$ constructed from the projected, traceless, symmetrized version of $D_{\!\mu} u_{\nu}$ vanishes as does the vector $E_{\mu}{-}T\partial_{\mu}\frac{\mu}{T}$. We note that both $\sigma^{\mu\nu}$ and $E_{\mu}{-}T\partial_{\mu}\frac{\mu}{T}$ contribute to dissipation, consistent with the claim that we are studying equilibrium states. 

By varying $W_1$ with respect to the metric and gauge field and decomposing according to \eqref{E:TJdecomp} and \eqref{E:expansion} we find
\begin{eqnarray}
&& \!\!\!\!\tilde{\pi}_{i}  {=} \gamma_{i} {=} \delta_{i} {=} 0, \quad
 \tilde{\phi}_{2} {=}\tilde{\gamma}_{1}{=} \dot{\tilde{\alpha}}_{2} {-} \tilde{\alpha}_{1}\,, \quad
 \tilde{\phi}_{1} {=} \tilde{\delta}_{1} {=} \dot{\tilde{\alpha}}_{1} \,, \label{E:PVresult1}
 \\ \nonumber
&& \!\!\!\!\tilde{\epsilon}_{1}  {=}T\tilde{\delta}_{2} {=}  T\tilde{\alpha}_{1}' {+} \mu \dot{\tilde{\alpha}}_{1} {-} \tilde{\alpha}_{1},\quad
  \tilde{\epsilon}_{2} {=} T\tilde{\gamma}_{2} {=} T\tilde{\alpha}_{2}' {+} \mu \dot{\tilde{\alpha}}_{2} {-} 2\tilde{\alpha}_{2}, 
\end{eqnarray}
where a prime denotes a derivative with respect to $T$ and a dot a derivative with respect to $\mu$. 

\begin{table}
\begin{tabular}{|l|c|c|}
\hline
 & $1$ & $2$ \\
\hline
pseudoscalars ($\tilde{s}_{i}$) & $ - \frac{1}{2\sqrt{-g}}\epsilon^{\mu\nu\rho}u_{\mu}F_{\nu\rho}$ & $ - \frac{1}{\sqrt{-g}}\epsilon^{\mu\nu\rho}u_{\mu}\partial_{\nu}u_{\rho}$  \\
vectors ($v_{i}$) & $E_{\mu}$ & $\Delta^{\mu\nu}\partial_{\nu} T$ \\
pseudovectors ($\tilde{v}_{i}$) & $\frac{1}{\sqrt{-g}}\epsilon^{\mu\nu\rho}u_{\nu}E_{\rho}$ & $\frac{1}{\sqrt{-g}}\epsilon^{\mu\nu\rho}u_{\nu}\partial_{\rho}T$ \\
\hline
\end{tabular}
\caption{\label{T:PVbasis}Independent first-order data for $2+1$ dimensional fluids. We have defined $E_{\mu}=F_{\mu\nu}u^{\nu}$.}
\end{table}

An analysis of parity-violating hydrodynamics in $2+1$ dimensions based on a local version of the second law of thermodynamics can be found in \cite{Jensen:2011xb}. Those results were presented in the Landau frame with
\begin{align}
\begin{split}
 {\cal  P} &= P - \tilde{\chi}_B \tilde{s}_1 - \tilde{\chi}_\Omega \tilde{s}_2 + \cdots \\
 \nu^\mu &= \chi_E v_1^\mu + \chi_T v_2^\mu  + \tilde{\chi}_E \tilde{v}_1^\mu + \tilde{\chi}_T \tilde{v}_2^\mu + \cdots
\end{split}
\end{align}
where the ellipsis denotes tensors which vanish in the equilibrium states under consideration. Matching the thermodynamic result (\ref{E:PVresult1}) to the Landau frame coefficients we find
\begin{align}
\begin{split}
\label{E:ToLandau}
	\tilde{\chi}_B &= \tilde{\pi}_1 - \dPde \tilde{\epsilon}_1 - \dPdr \tilde{\phi}_1\,, \quad
	\tilde{\chi}_E  = \tilde{\delta}_1 -R\tilde{\gamma}_1 \,,
	\\
	\tilde{\chi}_{\Omega} &= \tilde{\pi}_2 - \dPde \tilde{\epsilon}_2 - \dPdr \tilde{\phi}_2\,, \quad 
	\tilde{\chi}_T  = \tilde{\delta}_2 -R \tilde{\gamma}_2\,, 
	\\
	\chi_E &= \delta_1 - R \gamma_1\,, \quad 
	\chi_T  = \delta_2 - R\gamma_2\,,  
\end{split}
\end{align}
with $R = \rho/(\epsilon+P)$, where $\dPdr$ and $\dPde$ are evaluated at fixed $\epsilon$ and $\rho$ respectively. From \eqref{E:ToLandau} we find $\chi_E=\chi_T=0$ along with two relations among the four $\tilde{\chi}$'s. 
These relations are identical to those found in \cite{Jensen:2011xb} with $f_{\Omega}=0$, $\mathcal{M}_B = \tilde{\alpha}_1$ and $\mathcal{M}_{\Omega}=\tilde{\alpha}_2$. We emphasize that coefficients associated with tensor structures which vanish are undetermined by this method.

\emph{Example 3: Second-order hydrodynamics.}---In our final example, we consider a parity-preserving theory in $d$ spacetime dimensions without conserved $U(1)$ currents to second order in derivatives. There is only one zeroth order scalar, $s_{0,1}=T$. A computation similar to the one for the $2+1$ dimensional fluid implies that there are no first order scalars and  four second order scalars \footnote{As noted in \cite{Banerjee:2012iz}, the combination 
$\omega^{\mu\nu}\omega_{\mu\nu} + R_{\mu\nu} u^\mu u^\nu$ % equal to D_\mu u^\mu
is a total derivative, and thus in fact there are only three independent $\alpha_i$ parameters. One can
show that the results below are invariant under the shifts $\alpha_1 \rightarrow \alpha_1$, $\alpha_2 \rightarrow 0$, $\alpha_3 \rightarrow \alpha_3 - \alpha_2$
and $\alpha_4 \rightarrow \alpha_4 + T \alpha'_2$.}; see Table~\ref{T:2ndBasis}. Using similar notation to the previous example, we will drop the derivative index from the quantities in \eqref{E:Wderivative} and~\eqref{E:expansion}.
\begin{table}
\begin{tabular}{|l|c|c|c|c|}
\hline
 & $1$ & $2$ & $3$ & $4$ \\
\hline
scalars $(s_{i})$ & $R$ & $ u^{\mu}R_{\mu\nu}u^{\nu}$ & $\omega^{\mu\nu}\omega_{\nu\mu}$ & $a^{\mu}a_{\mu}$  \\
vectors $(v_{i})$ & $\Delta^{\mu\nu}R_{\nu\rho}u^{\rho}$ & $\omega^{\mu\nu}a_{\nu}$ & & \\
tensors $ (t_{i})$ & $R^{\langle\mu\nu\rangle }$ & $-u_{\rho} R^{\rho\langle\mu\nu\rangle \sigma}u_{\sigma}$ &  $\omega^{\langle \mu\rho}\omega_{\rho}^{\phantom{\rho}\nu\rangle }$ & $a^{\langle \mu}a^{\nu\rangle }$ \\
\hline
\end{tabular}
\caption{
	\label{T:2ndBasis}Independent second order data. The expressions for $a$ and $\omega$ are given by the inline expression following \eqref{E:Du}. $R^{\mu}_{\,\,\nu\rho\sigma}$ is the Riemann tensor and $R$ the Ricci scalar. Triangular brackets denote a projected traceless symmetrized tensor, $A_{\langle \mu \nu \rangle} = \frac{1}{2}\Delta_{\mu\rho}\Delta_{\nu\sigma}\left( A^{\rho\sigma}+A^{\sigma\rho}-\frac{2}{d-1}g^{\rho\sigma}\Delta_{\alpha\beta}A^{\alpha\beta}\right)$. }
\end{table}

Varying the generating functional $W_2$ defined in \eqref{E:Wderivative} with respect to the metric and expanding 
 the resulting energy-momentum tensor  according to \eqref{E:TJdecomp} and \eqref{E:expansion} we find for $d>3$,
\begin{align}
\begin{split}
\label{E:2ndO}
	\epsilon_{1} &= T\alpha_{1}'{-}\alpha_{1}\,, \qquad
	\epsilon_{2} = 2T(\alpha_{2}'{-}\alpha_{1}'){-}2\alpha_{1}{+}2\alpha_{4}\,, \\
	\epsilon_{3} &= T(\alpha_{3}'{+}\alpha_{2}'{-}2\alpha_{1}')+3(\alpha_{2}{-}\alpha_{3}){+}2\alpha_{4} \,, \\
	\epsilon_{4} & = T^2(2\alpha_{1}''{-}\alpha_{2}'') + 2T(2\alpha_{1}'{-}\alpha_{2}') {-} T\alpha_{4}' {-}\alpha_{4} \,, \\
	\pi_{1} &= \frac{d{-}3}{d{-}1} \alpha_{1}, \qquad  
	\pi_{2} = \frac{2(d{-}2)}{d{-}1}  T \alpha_{1}' - \frac{2}{d{-}1} \alpha_{1}, \\ 
	\pi_{3}  &= (\alpha_{3}-\alpha_{2})\frac{d{-}5}{d{-}1} + \frac{2(d{-}2)}{d{-}1} T\alpha_{1}', \\
	\pi_{4} &= (\alpha_{4} + T(\alpha_{2}'{-}2\alpha_{1}'))\frac{d{-}3}{d{-}1} - \frac{2(d{-}2)}{d{-}1} T^2 \alpha_{1}'', \\
	\theta_{1} &= -2 \alpha_{1}, \qquad 
	\theta_{3} = 4(\alpha_{2}-\alpha_{3})-2T \alpha_{1}',\\
	\theta_{2} &= -2T \alpha_{1}', \quad
	\theta_{4} = 2(T^2\alpha_{1}''{+}T(2\alpha_{1}'{-}\alpha_{2}'){-}\alpha_{4}), \\
	\gamma_{1} &= 2 (\alpha_{1}{+}\alpha_{2}{-}\alpha_{3}), \, \quad
	\gamma_{2} = -2 T  (\alpha_{1}'{+}\alpha_{2}'{-}\alpha_{3}') \,, 
\end{split}
\end{align}
where the  tensor structures $t_{i}$, $v_{i}$ and $s_{i}$ are given in Table \ref{T:2ndBasis}. We refer the reader to~\cite{Bhattacharyya:2012ex} for a comprehensive discussion of second order tensor structures. In three space-time dimensions~\eqref{E:2ndO} still describes the expansion~\eqref{E:expansion} of the energy-momentum tensor, but the list of tensors is overcomplete. In particular, the tensors $t_{3}$ and $t_{1}+t_{2}$ vanish so that only the combinations $\theta_1-\theta_2$ and $\theta_4$ appear in $\tau^{\mu\nu}$.

An analysis of the restrictions on response coefficients of $3+1$ dimensional systems to second order in the derivative expansion was carried out in \cite{Bhattacharyya:2012ex} (see also \cite{Romatschke:2009kr}). The results were presented in the Landau frame where
\begin{align}
\begin{split}
{\cal P} =& T\left(\zeta_2 s_1 + \zeta_3 s_2 + \xi_3 s_3 +\xi_4 s_4\right) + \cdots \\
 \tau^{\mu\nu} =& T\left(\kappa_1 t_1^{\mu\nu} + \kappa_2 t_2^{\mu\nu} +\lambda_3 t_3^{\mu\nu} + \lambda_4 t_4^{\mu\nu}\right) +\cdots.
\end{split}
\end{align}
The frame transformation from the thermodynamic frame to the Landau frame is given by $\delta u^{\mu} = q^{\mu}/(\epsilon+P)$ and $\delta T = (\epsilon - \E)/\epsilon'$. After carrying out this frame transformation, 
we find
\begin{align}
\begin{split}
T\kappa_1 &= \theta_1, \,\, T\kappa_2 = \theta_2, \,\, T\lambda_3 = \theta_3, \,\, T\lambda_4 = \theta_4, \\
T\zeta_2 &= \pi_1 - \dPde \epsilon_1, \quad T\zeta_3 = \pi_2 - \dPde \epsilon_2, \\
 T\xi_3 &= \pi_3 - \dPde \epsilon_3, \quad T\xi_4 = \pi_4 - \dPde\epsilon_4,
\end{split}
\end{align}
where $\dPde = \frac{s}{T}\frac{dT}{ds}$. Despite the fact that there are only four $\alpha_i$'s, one can verify that there are five relations between the eight coefficients above. These correspond precisely to the five conditions on the response coefficients found in~\cite{Bhattacharyya:2012ex}. The remaining seven transport coefficients are undetermined either by requiring the existence of an entropy current or by the variational method described in this Letter. 
The results for $d \neq 4$ are new; the variational method provides a simple alternative to the more onerous technique which uses the entropy current.

\emph{Discussion.}---In this work we have studied the implications of the existence of an equilibrium state on the hydrodynamic constitutive relations. We have shown, using three examples, how relations among response coefficients which are canonically derived using a local version of the second law of thermodynamics emerge from properties of the gauge- and diffeomorphism-invariant generating functional (\ref{E:Wgeneral}). 

Varying the generating functional with respect to the sources leads to zero frequency $n$-point Euclidean correlation functions. Since the dependence of the generating functional on the sources is known explicitly, there is no need to solve the equations of hydrodynamics in order to compute the correlators. This significantly reduces the complexity of the computation.

More importantly, the thermodynamic relations following from the generating functional combined with the inequalities imposed on dynamical transport coefficients from positivity of spectral functions appear to reproduce the entire suite of constraints implied by an entropy current with positive semi-definite divergence. Put differently, requiring the existence of a local entropy current with positive semi-definite divergence implies several constraints among the coefficients in the hydrodynamic equations. Some of these constraints appear in the form of equalities and others in the form of inequalities. 
As we have suggested in this Letter, the former constraints can be obtained by appealing to equilibrium thermodynamics. This suggestion may appear less surprising once we realize that equality-type constraints relate to dissipationless contributions to the energy-momentum tensor and charge current. It would be interesting to study whether all inequality-type constraints associated with dissipation follow in general from the positivity properties of even $n$-point functions.

While this work was in progress, we received an advance copy of Ref.~\cite{Banerjee:2012iz} which overlaps with the content of this Letter \footnote{Ref.~\cite{Banerjee:2012iz}, and the companion paper \cite{Jensen:2012jy} to the present work, also consider systems with anomalous currents.}. 
We thank the authors of \cite{Banerjee:2012iz} for informing us of their work  prior to publication.  We thank S.~Bhattacharyya, Z.~Komargodski, S.~Minwalla, and D.~T.~Son for useful discussions. KJ,~PK, and~AR were supported in part by NSERC, Canada. MK was supported by the US Department of Energy under contract number DE-FGO2-96ER40956. RM's work is partially supported by  European Union grants FP7-REGPOT-2008-1-CreteHEP
 Cosmo-228644 and PERG07-GA-2010-268246, as well as EU program ``Thalis" ESF/NSRF 2007-2013. AY~is a Landau fellow, supported in part by the Taub foundation and supported in part by the ISF under grant number 495/11 and by the BSF under grant number 2014350.

\bibliography{hwec-v2}

\begin{thebibliography}{23}%
\makeatletter
\providecommand \@ifxundefined [1]{%
 \@ifx{#1\undefined}
}%
\providecommand \@ifnum [1]{%
 \ifnum #1\expandafter \@firstoftwo
 \else \expandafter \@secondoftwo
 \fi
}%
\providecommand \@ifx [1]{%
 \ifx #1\expandafter \@firstoftwo
 \else \expandafter \@secondoftwo
 \fi
}%
\providecommand \natexlab [1]{#1}%
\providecommand \enquote  [1]{``#1''}%
\providecommand \bibnamefont  [1]{#1}%
\providecommand \bibfnamefont [1]{#1}%
\providecommand \citenamefont [1]{#1}%
\providecommand \href@noop [0]{\@secondoftwo}%
\providecommand \href [0]{\begingroup \@sanitize@url \@href}%
\providecommand \@href[1]{\@@startlink{#1}\@@href}%
\providecommand \@@href[1]{\endgroup#1\@@endlink}%
\providecommand \@sanitize@url [0]{\catcode `\\12\catcode `\$12\catcode
  `\&12\catcode `\#12\catcode `\^12\catcode `\_12\catcode `\%12\relax}%
\providecommand \@@startlink[1]{}%
\providecommand \@@endlink[0]{}%
\providecommand \url  [0]{\begingroup\@sanitize@url \@url }%
\providecommand \@url [1]{\endgroup\@href {#1}{\urlprefix }}%
\providecommand \urlprefix  [0]{URL }%
\providecommand \Eprint [0]{\href }%
\providecommand \doibase [0]{http://dx.doi.org/}%
\providecommand \selectlanguage [0]{\@gobble}%
\providecommand \bibinfo  [0]{\@secondoftwo}%
\providecommand \bibfield  [0]{\@secondoftwo}%
\providecommand \translation [1]{[#1]}%
\providecommand \BibitemOpen [0]{}%
\providecommand \bibitemStop [0]{}%
\providecommand \bibitemNoStop [0]{.\EOS\space}%
\providecommand \EOS [0]{\spacefactor3000\relax}%
\providecommand \BibitemShut  [1]{\csname bibitem#1\endcsname}%
\let\auto@bib@innerbib\@empty
%</preamble>
\bibitem [{\citenamefont {Landau}\ and\ \citenamefont {Lifshitz}(1987)}]{LL6}%
  \BibitemOpen
  \bibfield  {author} {\bibinfo {author} {\bibfnamefont {L.~D.}\ \bibnamefont
  {Landau}}\ and\ \bibinfo {author} {\bibfnamefont {E.~M.}\ \bibnamefont
  {Lifshitz}},\ }\href@noop {} {\emph {\bibinfo {title} {A Course in
  Theoretical Physics - Fluid Mechanics}}},\ Vol.~\bibinfo {volume} {6}\
  (\bibinfo  {publisher} {Pergamon},\ \bibinfo {year} {1987})\BibitemShut
  {NoStop}%
\bibitem [{\citenamefont {Bhattacharyya}\ \emph {et~al.}(2008)\citenamefont
  {Bhattacharyya} \emph {et~al.}}]{Bhattacharyya:2008jc}%
  \BibitemOpen
  \bibfield  {author} {\bibinfo {author} {\bibfnamefont {S.}~\bibnamefont
  {Bhattacharyya}} \emph {et~al.},\ }\href {\doibase
  10.1088/1126-6708/2008/02/045} {\bibfield  {journal} {\bibinfo  {journal}
  {JHEP}\ }\textbf {\bibinfo {volume} {02}},\ \bibinfo {pages} {045} (\bibinfo
  {year} {2008})}\BibitemShut {NoStop}%
%%CITATION = 0712.2456;%%
\bibitem [{\citenamefont {Baier}\ \emph {et~al.}(2008)\citenamefont {Baier}
  \emph {et~al.}}]{Baier:2007ix}%
  \BibitemOpen
  \bibfield  {author} {\bibinfo {author} {\bibfnamefont {R.}~\bibnamefont
  {Baier}} \emph {et~al.},\ }\href {\doibase 10.1088/1126-6708/2008/04/100}
  {\bibfield  {journal} {\bibinfo  {journal} {JHEP}\ }\textbf {\bibinfo
  {volume} {04}},\ \bibinfo {pages} {100} (\bibinfo {year} {2008})}\BibitemShut
  {NoStop}%
%%CITATION = 0712.2451;%%
\bibitem [{Note1()}]{Note1}%
  \BibitemOpen
  \bibinfo {note} {There exist several distinct proposals in the literature
  which define a dynamical effective action for non-dissipative hydrodynamics
  \cite
  {Schutz:1970my,*Dixon:1979,*Brown:1992kc,*Son:2002zn,*Jackiw:2004nm,*Dubovsky:2005xd,*Dubovsky:2011sj}.}\BibitemShut
  {Stop}%
\bibitem [{Note2()}]{Note2}%
  \BibitemOpen
  \bibinfo {note} {Some readers may be familiar with the Euclidean expression
  $L= \DOTSI \intop \ilimits@ _0^\beta d\tau \protect \sqrt {g_{\tau \tau }}$
  which can be obtained from $L=\beta \protect \sqrt {-V^2}$ by setting $V^\mu
  = (1,{\protect \bf 0})$ and rotating to Euclidean time.}\BibitemShut {Stop}%
\bibitem [{\citenamefont {Son}(2001)}]{Son:2000ht}%
  \BibitemOpen
  \bibfield  {author} {\bibinfo {author} {\bibfnamefont {D.}~\bibnamefont
  {Son}},\ }\href@noop {} {\bibfield  {journal} {\bibinfo  {journal}
  {Int.J.Mod.Phys.A}\ }\textbf {\bibinfo {volume} {16 S1C}},\ \bibinfo {pages}
  {1284} (\bibinfo {year} {2001})},\ \Eprint
  {http://arxiv.org/abs/hep-ph/0011246} {arXiv:hep-ph/0011246 [hep-ph]}
  \BibitemShut {NoStop}%
%%CITATION = HEP-PH/0011246;%%
\bibitem [{\citenamefont {Herzog}\ \emph {et~al.}(2009)\citenamefont {Herzog},
  \citenamefont {Kovtun},\ and\ \citenamefont {Son}}]{Herzog:2008he}%
  \BibitemOpen
  \bibfield  {author} {\bibinfo {author} {\bibfnamefont {C.}~\bibnamefont
  {Herzog}}, \bibinfo {author} {\bibfnamefont {P.}~\bibnamefont {Kovtun}}, \
  and\ \bibinfo {author} {\bibfnamefont {D.}~\bibnamefont {Son}},\ }\href
  {\doibase 10.1103/PhysRevD.79.066002} {\bibfield  {journal} {\bibinfo
  {journal} {Phys.Rev.D}\ }\textbf {\bibinfo {volume} {79}},\ \bibinfo {pages}
  {066002} (\bibinfo {year} {2009})}\BibitemShut {NoStop}%
%%CITATION = ARXIV:0809.4870;%%
\bibitem [{Note3()}]{Note3}%
  \BibitemOpen
  \bibinfo {note} {For an ideal superfluid, a sufficient condition for the
  conservation laws (\ref {E:conservation}) to hold in equilibrium is that
  $D_\mu (f\xi ^\mu )=0$, the equation of motion for $\phi $. It would be
  interesting to identify the geometric criteria which will determine
  equilibrium configurations of superfluids in the presence of an external
  metric and gauge field. We thank S. Bhattacharyya for emphasizing this
  point.}\BibitemShut {Stop}%
\bibitem [{\citenamefont {Bhattacharya}\ \emph {et~al.}(2011)\citenamefont
  {Bhattacharya} \emph {et~al.}}]{Bhattacharya:2011tra}%
  \BibitemOpen
  \bibfield  {author} {\bibinfo {author} {\bibfnamefont {J.}~\bibnamefont
  {Bhattacharya}} \emph {et~al.},\ }\href@noop {} {\  (\bibinfo {year}
  {2011})},\ \Eprint {http://arxiv.org/abs/1105.3733} {arXiv:1105.3733}
  \BibitemShut {NoStop}%
%%CITATION = ARXIV:1105.3733;%%
\bibitem [{\citenamefont {Jensen}\ \emph {et~al.}(2011)\citenamefont {Jensen}
  \emph {et~al.}}]{Jensen:2011xb}%
  \BibitemOpen
  \bibfield  {author} {\bibinfo {author} {\bibfnamefont {K.}~\bibnamefont
  {Jensen}} \emph {et~al.},\ }\href@noop {} {\  (\bibinfo {year} {2011})},\
  \Eprint {http://arxiv.org/abs/1112.4498} {arXiv:1112.4498} \BibitemShut
  {NoStop}%
%%CITATION = ARXIV:1112.4498;%%
\bibitem [{Note4()}]{Note4}%
  \BibitemOpen
  \bibinfo {note} {As noted in \cite {Banerjee:2012iz}, the combination $\omega
  ^{\mu \nu }\omega _{\mu \nu } + R_{\mu \nu } u^\mu u^\nu $ is a total
  derivative, and thus in fact there are only three independent $\alpha _i$
  parameters. One can show that the results below are invariant under the
  shifts $\alpha _1 \rightarrow \alpha _1$, $\alpha _2 \rightarrow 0$, $\alpha
  _3 \rightarrow \alpha _3 - \alpha _2$ and $\alpha _4 \rightarrow \alpha _4 +
  T \alpha '_2$.}\BibitemShut {Stop}%
\bibitem [{\citenamefont {Bhattacharyya}(2012)}]{Bhattacharyya:2012ex}%
  \BibitemOpen
  \bibfield  {author} {\bibinfo {author} {\bibfnamefont {S.}~\bibnamefont
  {Bhattacharyya}},\ }\href@noop {} {\  (\bibinfo {year} {2012})},\ \Eprint
  {http://arxiv.org/abs/1201.4654} {arXiv:1201.4654} \BibitemShut {NoStop}%
%%CITATION = ARXIV:1201.4654;%%
\bibitem [{\citenamefont {Romatschke}(2010)}]{Romatschke:2009kr}%
  \BibitemOpen
  \bibfield  {author} {\bibinfo {author} {\bibfnamefont {P.}~\bibnamefont
  {Romatschke}},\ }\href {\doibase 10.1088/0264-9381/27/2/025006} {\bibfield
  {journal} {\bibinfo  {journal} {Class.Quant.Grav.}\ }\textbf {\bibinfo
  {volume} {27}},\ \bibinfo {pages} {025006} (\bibinfo {year} {2010})},\
  \Eprint {http://arxiv.org/abs/0906.4787} {arXiv:0906.4787 [hep-th]}
  \BibitemShut {NoStop}%
%%CITATION = ARXIV:0906.4787;%%
\bibitem [{\citenamefont {Banerjee}\ \emph {et~al.}(2012)\citenamefont
  {Banerjee} \emph {et~al.}}]{Banerjee:2012iz}%
  \BibitemOpen
  \bibfield  {author} {\bibinfo {author} {\bibfnamefont {N.}~\bibnamefont
  {Banerjee}} \emph {et~al.},\ }\href@noop {} {\  (\bibinfo {year} {2012})},\
  \Eprint {http://arxiv.org/abs/1203.3544} {arXiv:1203.3544 [hep-th]}
  \BibitemShut {NoStop}%
%%CITATION = ARXIV:1203.3544;%%
\bibitem [{Note5()}]{Note5}%
  \BibitemOpen
  \bibinfo {note} {Ref.~\cite {Banerjee:2012iz}, and the companion paper \cite
  {Jensen:2012jy} to the present work, also consider systems with anomalous
  currents.}\BibitemShut {Stop}%
\bibitem [{\citenamefont {Schutz}(1970)}]{Schutz:1970my}%
  \BibitemOpen
  \bibfield  {author} {\bibinfo {author} {\bibfnamefont {B.~F.}\ \bibnamefont
  {Schutz}},\ }\href {\doibase 10.1103/PhysRevD.2.2762} {\bibfield  {journal}
  {\bibinfo  {journal} {Phys.Rev.D}\ }\textbf {\bibinfo {volume} {2}},\
  \bibinfo {pages} {2762} (\bibinfo {year} {1970})}\BibitemShut {NoStop}%
%%CITATION = PHRVA,D2,2762;%%
\bibitem [{\citenamefont {Dixon}(1979)}]{Dixon:1979}%
  \BibitemOpen
  \bibfield  {author} {\bibinfo {author} {\bibfnamefont {W.~G.}\ \bibnamefont
  {Dixon}},\ }\href@noop {} {\bibfield  {journal} {\bibinfo  {journal} {Arch.
  Rational Mech. Anal.}\ }\textbf {\bibinfo {volume} {69}},\ \bibinfo {pages}
  {293} (\bibinfo {year} {1979})}\BibitemShut {NoStop}%
\bibitem [{\citenamefont {Brown}(1993)}]{Brown:1992kc}%
  \BibitemOpen
  \bibfield  {author} {\bibinfo {author} {\bibfnamefont {J.}~\bibnamefont
  {Brown}},\ }\href {\doibase 10.1088/0264-9381/10/8/017} {\bibfield  {journal}
  {\bibinfo  {journal} {Class.Quant.Grav.}\ }\textbf {\bibinfo {volume} {10}},\
  \bibinfo {pages} {1579} (\bibinfo {year} {1993})}\BibitemShut {NoStop}%
%%CITATION = GR-QC/9304026;%%
\bibitem [{\citenamefont {Son}(2002)}]{Son:2002zn}%
  \BibitemOpen
  \bibfield  {author} {\bibinfo {author} {\bibfnamefont {D.}~\bibnamefont
  {Son}},\ }\href@noop {} {\  (\bibinfo {year} {2002})},\ \Eprint
  {http://arxiv.org/abs/hep-ph/0204199} {arXiv:hep-ph/0204199 [hep-ph]}
  \BibitemShut {NoStop}%
%%CITATION = HEP-PH/0204199;%%
\bibitem [{\citenamefont {Jackiw}\ \emph {et~al.}(2004)\citenamefont {Jackiw}
  \emph {et~al.}}]{Jackiw:2004nm}%
  \BibitemOpen
  \bibfield  {author} {\bibinfo {author} {\bibfnamefont {R.}~\bibnamefont
  {Jackiw}} \emph {et~al.},\ }\href {\doibase 10.1088/0305-4470/37/42/R01}
  {\bibfield  {journal} {\bibinfo  {journal} {J.Phys.A}\ }\textbf {\bibinfo
  {volume} {37}},\ \bibinfo {pages} {R327} (\bibinfo {year}
  {2004})}\BibitemShut {NoStop}%
%%CITATION = HEP-PH/0407101;%%
\bibitem [{\citenamefont {Dubovsky}\ \emph {et~al.}(2006)\citenamefont
  {Dubovsky} \emph {et~al.}}]{Dubovsky:2005xd}%
  \BibitemOpen
  \bibfield  {author} {\bibinfo {author} {\bibfnamefont {S.}~\bibnamefont
  {Dubovsky}} \emph {et~al.},\ }\href {\doibase 10.1088/1126-6708/2006/03/025}
  {\bibfield  {journal} {\bibinfo  {journal} {JHEP}\ }\textbf {\bibinfo
  {volume} {0603}},\ \bibinfo {pages} {025} (\bibinfo {year}
  {2006})}\BibitemShut {NoStop}%
%%CITATION = HEP-TH/0512260;%%
\bibitem [{\citenamefont {Dubovsky~{\it et al.}}(2011)}]{Dubovsky:2011sj}%
  \BibitemOpen
  \bibfield  {author} {\bibinfo {author} {\bibfnamefont {S.}~\bibnamefont
  {Dubovsky~{\it et al.}}},\ }\href@noop {} {\  (\bibinfo {year} {2011})},\
  \Eprint {http://arxiv.org/abs/1107.0731} {arXiv:1107.0731 [hep-th]}
  \BibitemShut {NoStop}%
%%CITATION = ARXIV:1107.0731;%%
\bibitem [{\citenamefont {Jensen}(2012)}]{Jensen:2012jy}%
  \BibitemOpen
  \bibfield  {author} {\bibinfo {author} {\bibfnamefont {K.}~\bibnamefont
  {Jensen}},\ }\href@noop {} {\  (\bibinfo {year} {2012})},\ \Eprint
  {http://arxiv.org/abs/1203.3599} {arXiv:1203.3599 [hep-th]} \BibitemShut
  {NoStop}%
%%CITATION = ARXIV:1203.3599;%%
\end{thebibliography}%

\end{document}